\documentclass[aps,twocolumn,preprintnumbers,showpacs,showkeys]{revtex4}
\usepackage{epsfig}
\usepackage{amssymb,amsmath,amsfonts,amsthm,graphicx}
\graphicspath{{./Figures/}}                 % location for color  fig.
\newcommand{\noi}{\noindent}
\newcommand{\beq}{\begin{equation}}
\newcommand{\eeq}{\end{equation}}
\newcommand{\bea}{\begin{array}}
\newcommand{\eea}{\end{array}}
\newcommand{\beqa}{\begin{eqnarray}}
\newcommand{\eeqa}{\end{eqnarray}}
\newcommand{\Fig}[1]{Fig.~\ref{#1}}

\newcommand{\Sec}[1]{Section~\ref{#1}}
\newcommand{\Eq}[1]{Eq.~(\ref{#1})}

\def\beqa{\begin{eqnarray}}
\def\eeqa{\end{eqnarray}}
\def\pl{{{\cal P}_\infty}}

\def\Tr{{\rm Tr}}

\begin{document}
%% VB preprint number corrected
\preprint{ITEP-LAT/2015-18,~HU-EP-15/59}

\title{Dyons near the transition temperature in lattice QCD \\}

\author{V.~G.~Bornyakov}
%\email[]{bornvit@gmail.com}
\affiliation{Institute for High Energy Physics NRC ``Kurchatov Institute'',
142281 Protvino, Russia, \\
Institute of Theoretical and Experimental Physics, 117259 Moscow, 
Russia \\
School of Biomedicine, Far East Federal University, 690950 Vladivostok, 
Russia}

\author{E.-M. Ilgenfritz}
%\email[]{Michael.Ilgenfritz@sunse.jinr.ru}
\affiliation{Joint Institute for Nuclear Research, BLTP, 141980 Dubna, Russia}

\author{B.~V.~Martemyanov}
%\email[]{martemja@itep.ru}
\affiliation{Institute of Theoretical and Experimental Physics, 117259 Moscow, 
Russia \\
National Research Nuclear University MEPhI, 115409, Moscow, Russia \\
Moscow Institute of Physics and Technology, 141700, Dolgoprudny, Moscow Region,
Russia}

\author{\fbox{M.~M\"uller-Preussker}}
%\email[]{mmp@physik.hu-berlin.de}
\affiliation{Humboldt-Universit\"at zu Berlin, Institut f\"ur Physik,
  12489 Berlin, Germany}

\date{2015}
\date{\today}
%------------------------------------------------------------------------------
\begin{abstract}
 We study the topological structure of QCD by cluster analysis. The fermionic topological
charge density is constructed from low lying modes of overlap Dirac operator for three
types of temporal boundary conditions for fermion field. This gives the possibility to mark
all three dyon constitutents of KvBLL caloron in gluonic fields. The gluonic topological
charge density is appearing in the process of overimproved gradient flow process stopped
at the moment when it  maximally matches the fermionic topological charge density. This 
corresponds to the smearing of gluonic fields up to the scale set by dyon size.
The time-like Abelian monopoles and specific  KvBLL pattern of Polyakov line correlate
with topological clusters.
\end{abstract}



\maketitle

\section{Introduction}
%---------------------
\label{sec:introduction}
Two basic properties of QCD are confinement (of quarks and gluons) and the
spontaneous breaking of chiral symmetry at low temperature and density.
Both properties are believed to be intimately connected with each other and
to originate from a certain complex structure of the QCD vacuum state, the
simplest manifestations of which being condensates of gluon and quark fields.
The field fluctuations contributing to these condensates are, however,
space-time and scale dependent. One of the aims of Lattice Gauge Theory
is to reveal the corresponding structures. One school of thought claims
that - at the infrared scale - the origin of both mechanisms can be
traced back to semiclassical objects of QCD. These objects either disappear 
or change their properties at high temperature, where the quark-gluon 
plasma phase appears.

Today it is commonplace to say
that the instanton mechanism is able to explain  chiral symmetry breaking while
it fails to provide a mechanism for confinement. Without further 
sophistications this is definitely correct for the instanton gas or liquid. 
Constituent dyons of Kraan-van Baal-Lee-Lu (KvBLL) calorons~\cite{Kraan:1998pm,
Kraan:1998sn,Lee:1998bb},
however, are as good as instantons in explaining chiral symmetry breaking. 
Calorons with their dyon ``substructure'' give some room to reproduce certain
features of confinement (Polyakov loop correlators, spatial string tension,
vortex and/or monopole percolation) which was the reason why people expected
for decades that ``instanton quarks'' might solve the confinement problem. 
Moreover, dyons when considered as rarefied gas, either without 
interaction or with Coulomb-like interaction, give confining behavior for 
space-like Wilson loops and for correlators of Polyakov loops. The history 
of this idea ranges from the 70's to the recent past~\cite{Polyakov:1976fu,
Martemyanov:1997ks,Gerhold:2006sk,Diakonov:2007nv,Bruckmann:2011yd}.

The modelling of dyon ensembles with interaction has got even more attraction recently
~\cite{Shuryak:2011aa,Faccioli:2013ja,Larsen:2014yya,Liu:2015ufa,Liu:2015jsa,Larsen:2015vaa}.
Therefore, it is of some interest to search for dyons in thermal
Monte Carlo configurations (representing lattice gauge fields at different
temperatures) in order to assess the relevance of these models and in order
to eventually observe dyons clustering into caloron-like quasiclassical
configurations.

The caloron with nontrivial holonomy~\cite{Kraan:1998pm,Kraan:1998sn,Lee:1998bb}
has the remarkable property that the single zero mode of the Dirac operator
is able to locate on distinct constituent 
dyons~\cite{GarciaPerez:1999ux,Chernodub:1999wg},
depending on the temporal boundary codition (b.c.) applied to the Dirac 
operator if it possesses improved chiral properties.
  
Inspired by the KvBLL solutions, for a subset of thermal lattice configurations
of fixed total topological charge $Q=\pm 1$ (created below and above 
$T_{\rm dec}$) the change of the single zero mode's localization with the 
change of b.c. was observed by Gattringer et al. 
~\cite{Gattringer:2002wh,Gattringer:2002tg} and interpreted in the caloron
picture, ignoring other topological features of these configurations. 
In the case of $SU(2)$ and $SU(3)$ lattice gauge theory it has been seen 
that this property of mobility (and changing degree of localization) is shared 
also by a band of near-zero modes of the overlap Dirac 
operator~\cite{Bornyakov:2007fm,Bornyakov:2008im,Ilgenfritz:2013oda,Bornyakov:2014esa}. 

Thus, not only the set of zero modes reflecting the total topological charge
of the gluonic field, but the band of low lying modes of the overlap Dirac 
operator identified with different boundary conditions can be used as an 
effective tool to detect distinct topological objects. This direct insight 
in Monte Carlo configurations of lattice gauge fields (without cooling or 
smearing) is restricted, however, to a corresponding scale set by the 
eigenvalues. In the present paper we will see to what extent this technique 
leads us to the three dyons making up one ``instanton'' (caloron) of QCD.

In \Sec{sec:definitions2} we introduce the lattice set-up in which the
ensembles of gauge fields which we are going to analyze  have been generated 
in lattice QCD with $N_f=2$ dynamical flavors.
In \Sec{sec:topological_observables} we define all the topologically relevant 
lattice observables employed lateron for the analysis.
Then, in \Sec{sec:ensemble_results} the concrete case of $N_f=2$ lattice QCD 
is considered at two temperatures, at the crossover $T_\chi$ (ensemble I)
and at $T = 1.06 T_\chi$ (ensemble II).
%% VB changed the next sentence
%We constryct the fermionic topological charge density with the help
%of a set of low-lying Dirac eigenmodes for all three different fermionic
%boundary conditions. 
The fermionic topological charge density was constructed 
with the help of low lying modes of overlap Dirac operator computed 
for three types of temporal boundary conditions.
%% VB end

This is the point where two new ideas compared to the previous papers
(dealing with $SU(3)$ gluodynamics) are introduced.
(i) Again, in addition to antiperiodic boundary conditions for the overlap
fermions, two other boundary conditions are employed in order to construct
topological densities for each type of boundary conditions. Here, however,
not the number of pairs of non-zero modes is fixed for all boundary conditions,
but the eigenvalue cutoff is fixed. We consider two cutoffs at two temperatures.
(ii) In order to characterize the configurations by the gluonic topological
charge density, the technique of gradient flow has replaced the procedure of
cooling that was previously in use. The stopping criterion is now set by the
maximal approximation of the fermionic topological charge density by the
gluonic one. This is how the gluonic topological density depends on the
chosen cutoff scale.

It will turn out that in this way all UV fluctuations are removed, up to 
the size of dyons that we are searching for.

Finally, the properties of the clusters of the three fermionic topological 
densities under consideration are studied as well as their correlations
among each other, correlations to the local holonomy and to the Abelian 
monopoles from the Maximally Abelian Gauge construction.

In \Sec{sec:conclusions} we shall draw our conclusions.

%------------------------------------------------------------------------------
\section{Lattice Setting for the Thermal Ensembles}
%--------------------------------------------------
\label{sec:definitions2}

We continue our study of topological objects in $SU(3)$ gauge theory in
the case of QCD (with dynamical quarks). We have analyzed gauge field 
configurations generated with the Wilson gauge action $S_W$ and $N_f=2$ 
dynamical flavors of nonperturbatively $O(a)$ improved Wilson fermions 
(clover fermions).
The configurations had been produced long ago by the DIK 
collaboration~\cite{Bornyakov:2004ii,Bornyakov:2009qh} using the ``Berlin QCD'' code 
(BQCD)~\cite{Nakamura:2010qh}. The improvement coefficient $c_{SW}$ had 
been determined nonperturbatively \cite{Jansen:1998mx}.
The lattice spacing and pion mass had been determined by interpolation 
of $T=0$ results obtained by QCDSF collaboration \cite{Gockeler:2006jt}. 
We have analysed configurations produced on lattices with temporal extent 
$L_\tau=8$ and spatial sizes $L_s=16$ (ensemble I of 50 configurations at 
$T=T_\chi$) and 
$L_\tau=8$ and $L_s=24$ (ensemble II of 50 configurations at $T=1.06T_\chi$).
The DIK collaboration had scanned the temperature $T$ at fixed $\beta$-value
by changing the Wilson fermion hopping parameter $\kappa$. In other words,
the quark mass was not kept constant. 
%% VB restored T_\chi
The chiral crossover temperature 
$T_\chi \approx 230$ MeV was determined 
in 
Refs. \cite{Bornyakov:2004ii,Bornyakov:2009qh} at a corresponding pion 
mass 
value of $O(1~\mathrm{GeV})$. 
%The pion mass for the chosen  ensembles is of $O(1~\mathrm{GeV})$.

In Ref.\cite{Bornyakov:2013iva} we have investigated the $T$-dependence of the
topological susceptibility throughout the interval
$[0.85 T_\chi, 1.26 T_\chi]$ by overimproved cooling applied to
ensembles of 500 or 200 configurations. We have there
confronted the $T$-dependence with the case of pure $SU(3)$ Yang-Mills
theory (with Wilson action).

For the two temperatures $T=T_\chi$ and $T=1.06 T_\chi$, that we are going
to reinvestigate here with overlap fermions with respect to details of
the space-time topological structure (however for a smaller sub-ensemble
of 50 configurations each), we recall the topological susceptibility we
have found in Ref.\cite{Bornyakov:2013iva}:
\\ensemble I
(500 configurations, overimproved cooling),
$\chi_{\rm top} = (0.6\pm 0.05) T_\chi^4$;
\\ensemble II
(200 configurations, overimproved cooling),
$\chi_{\rm top} = (0.3\pm 0.03) T_\chi^4$.

\section{Topologically relevant observables}
%-------------------------------------------
\label{sec:topological_observables}

%% VB changed this  subsection 
We use in our analysis the following instruments (observables):
\begin{itemize}
\item  local holonomy and its trace (the Polyakov loop),
\item improved gluonic topological charge,
\item Abelian monopoles revealed by Abelian projection after 
transforming gauge field to the Maximal Abelian Gauge (MAG).
\end{itemize}
All these quantities are computed after gradient flow (in close 
correspondence to over-improved cooling).
We also use 
\begin{itemize}
\item the fermionic topological charge density (and its UV-filtered
version) including its dependence on the temporal boundary conditions
imposed on overlap fermions.
\end{itemize}

The importance and usefulness of
the finite-temperature holonomy (considered globally to distinguish the
phases of the theory and locally to distinguish the dyonic constituents 
or ``instanton quarks'') for the topological structure studies 
was recognized only through the discovery of 
the KvBLL-caloron solutions \cite{Kraan:1998pm,Kraan:1998sn,Lee:1998bb}.

%--------------------
\subsection{Holonomy}
\label{sec:holonomy}

%Let us begin with the local holonomy and its eigenvalues.
The local holonomy is defined as a product of timelike links
\beq
P(\vec{x}) = \prod_{x_0=1}^{N_\tau} U_0(\vec{x},x_0)
\label{eq:localholonomy}
\eeq
$P(\vec{x}) $ has eigenvalues
\beq
\lambda_k(\vec{x})=\exp\left(i 2\pi \mu_k(\vec{x})\right)\,.
\label{eq:localholonomy_evs}
\eeq

The positions in space of the dyon constituents
of KvBLL calorons are determined by the condition
that two of these eigenvalues coincide 
(cf. Appendix in \cite{Ilgenfritz:2013oda}). We  
use this property to localize (anti)dyons in 
unsmoothed lattice field configurations.

The asymptotic holonomy of KvBLL calorons
(after a suitable constant gauge transformation)
\beq
\pl \equiv \lim_{|\vec x|\rightarrow\infty} P(\vec{x}) =
\exp[2\pi i\,{\rm diag}(\mu_1,\mu_2,\mu_3)],
\eeq
is characterized by the eigenphases, three real and ordered numbers
$\mu_1\leq\ldots\leq\mu_3\leq\mu_{4}\!\equiv\!1\!+\!\mu_1$ fulfilling
$~\mu_1 + \mu_2 + \mu_3 \!=\!0$. The set of eigenphases
eventually determines the masses of well-separated dyon constituents 
via $8\pi^2\nu_m$, where $\nu_m\!\equiv\!\mu_{m+1}\!-\!\mu_m$
(cf. Appendix in \cite{Ilgenfritz:2013oda}).

The trace of $P(\vec{x})$ is the gauge invariant
complex-valued Polyakov loop
\beq
L(\vec{x}) = \frac{1}{3} \Tr~P(\vec{x}) \; .
\eeq
Its value can be represented as a point in 
the Weyl plot in the complex plane, see Fig.~ref{plincl} 
in Section~\ref{sec:ensemble_results}.

For $SU(3)$, when two eigenvalues of the
local holonomy are equal, respective Polyakov loop is represented
by the point in the periphery of the Weyl plot. If all three eigenvalues
coincide, the holonomy is an element of the center group
\beq
P(\vec{x}) = z_i \cdot I  \; ,
\eeq
where  $$z_i \in \{1,\exp{(2\pi i/3)},\exp{(-2\pi i/3)}\}.$$

The expectation value of the spatially averaged Polyakov loop ($V_3$ is the spatial volume) :
\beq
\overline L = \frac{1}{V_3}~\sum_{\vec{x}} L(\vec{x}) \; .
\eeq
is an order parameter of the 
deconfinement transition in pure Yang-Mills gauge theory, signalling  the
breaking of the center symmetry. In presence of dynamical fermions,
this symmetry is -- at best -- only approximate, even at low temperature.

%------------------------------------------------
%% VB next subsection was substantially reformulated and reduced
\subsection{MAG and Abelian monopoles definitions}
\label{sec:mag-monopolesa}
We use the definition of MAG introduced for lattice $SU(N)$ theory in 
\cite{Kronfeld:1987vd} and later specified for the $SU(3)$ group in 
\cite{Brandstater:1991sn}. The MAG is fixed by maximizing the functional
\beqa
\label{FU}
F[U] = \frac{1}{12\,V}~
 \sum_{x,\mu}~\left[ |(U_\mu(x))_{11}|^2 +|(U_\mu(x))_{22}|^2 \right. \nonumber \\
            \left. +|(U_\mu(x))_{33}|^2 \right]
\eeqa
with respect to local gauge transformations $g$ of the lattice gauge field,
\beq
U_\mu(x) \to U^g_\mu(x) = g(x)^\dagger U_\mu(x) g(x+\hat \mu)\,.
\eeq
Alternative definitions of the MAG condition for the $SU(3)$ group were
introduced in \cite{Tucker:2001tt} and were studied further in
\cite{Bonati:2013bga}.
To maximize the functional eq.~(\ref{FU}) we use the simulated annealing algorithm which
was found very much effective to fight the problem of Gribov copies \cite{Bali:1996dm}. For $SU(3)$ gauge 
group it was first used in \cite{Bornyakov:2001qw}, see also \cite{Bornyakov:2003vx} for details of its 
implementation to the case of the $SU(3)$ gauge group. 
After fixing to MAG the Abelian fields $u_\mu(x) \in U(1) \times U(1)$ are determined as a result of 
the Abelian projection described in \cite{Kronfeld:1987vd}.

The monopole currents $j^{(a)}_\mu(^*x)$ are defined \cite{Kronfeld:1987vd} on links of the dual lattice and 
satisfy the current conservation law for every $a$ separately:
\beq
\sum_{\mu}\partial^-_\mu j^{(a)}_\mu(s) = 0 \,, \,\, a=1,2,3 \,.
\eeq
Additionally 
\beq
\sum_{a=1}^3 j^{(a)}_\mu(x) = 0 \, ,
\eeq
i.e. only two currents are independent.

%-----------------------------------------------------------------------------
%% VB this subsection was also changed 
\subsection{The gluonic definition of the topological density}
\label{sec:topdensity}

The definition of topological charge density is
\begin{equation}
q(x) =  \frac{1}{16 \pi^2} \Tr (F_{\mu\nu}(x)\,\tilde{F}_{\mu\nu}(x))
\label{eq:qdensity}
\end{equation}
were
\begin{equation}
\tilde{F}_{\mu\nu}(x)=
 \frac{1}{2}~\epsilon_{\mu\nu\lambda\sigma}~F_{\lambda\sigma}(x)\,.
\end{equation}
The lattice gluonic topological charge density 
uses field strength definition of $F_{\mu\nu}(x)$
as a ``clover'' average over the traceless antihermitean part of all 
four plaquettes within the $\mu\nu$ plane with sidelength $n=1$  
placed around a site $x$ while kept untraced in that site $x$.
The improved topological charge density ~\cite{BilsonThompson:2002jk}
extends this construction to quadratic loops of sizes $n= 2, 3$ 
which are added with appropriate weights. 
The improved topological charge and corresponding  action 
(in units of the one-instanton action $S_{\rm inst}$) 
are then defined as 
%lattice sums over the improved field strength tensor  squared
\begin{eqnarray}
Q_{\rm glue} &=& \sum_x \Tr (F_{\mu\nu} (x)~\tilde{F}_{\mu\nu} (x)) / (16 \pi^2)\, , 
\label{gluonic_Q} \\
S/ S_{\rm inst} & = & \sum_x \Tr (F_{\mu\nu} (x)~F_{\mu\nu} (x)) / (16 \pi^2)\, .
\label{gluonic_S}
\end{eqnarray}

%-----------------------------------------------------------------------------
\subsection{Over-improved gradient flow}
\label{sec:cooling}

Gradient flow is an advanced method to remove quantum fluctuations
up to a certain ``diffusion'' scale from given lattice field
configurations created in the course of Monte Carlo simulations
~\footnote{In contrast to gradient flow, cooling has no adaptable
``flow time step'' $\Delta \tau$ and consists of sequential minimizations 
with respect to single links sweeped through the lattice. In the case 
of $SU(3)$ this requires a non-analytical operation called ``projection 
to the group''.}.
The gradient flow effectively results also in a minimization of the action 
in the ``direction'' of steepest descent in configuration 
space~\cite{Luscher:2009eq,Luscher:2010iy,Luscher:2011bx}. Proposed 
by L\"uscher for the Wilson (one-plaquette) action, the gradient flow 
can be defined with respect to different gluonic actions, of which
Wilson flow realizes the simplest case. In our case, we propose to use 
the gradient flow with respect of an action of the form
\begin{eqnarray}
S(\epsilon) &=& \sum_{x,\mu\nu}  \frac{4-\epsilon}{3}
            \mbox{Re~Tr}~\left( 1 - U_{x,\mu\nu} \right) \nonumber \\
            &+& \sum_{x,\mu\nu}  \frac{1-\epsilon}{48}
            \mbox{Re~Tr}~\left( 1 - U^{2\times2}_{x,\mu\nu} \right) \; ,
\end{eqnarray} which reduces to the Wilson action in the case $\epsilon=1$. 
The so-called over-improved action \cite{GarciaPerez:1993ki} corresponds 
to $\epsilon=-1$. 
Expanding in powers of lattice spacing $a$, one finds that the lattice action
includes now higher dimension operators:
\begin{equation}
S(\epsilon) = \sum_{x,\mu\nu} a^4  \Tr \left[  \frac{1}{2} F^2_{\mu \nu}(x)
- \frac{\epsilon a^2}{12} \left(D_\mu F_{\mu \nu}(x) \right)^2 \right]+O(a^8).
\end{equation}
For a discretized continuum instanton of size $\rho$ this provides
corrections of order $a/\rho$:
\begin{equation}
S(\epsilon) = 8 \pi^2 \left[ 1 - \frac{\epsilon}{5} \left(\frac{a}{\rho}\right)^2
+ {\cal O}\left(\left[\frac{a}{\rho}\right]^4\right) \right]
\end{equation}
suggesting that under cooling $\rho$ will decrease for $\epsilon > 0$
and increase for $\epsilon < 0$.
The inversion of lattice artefacts relative to the Wilson case makes
topological lumps stable against the process of gradient flow.

\begin{figure}[htb]
\centering 
\includegraphics[width=0.41\textwidth]{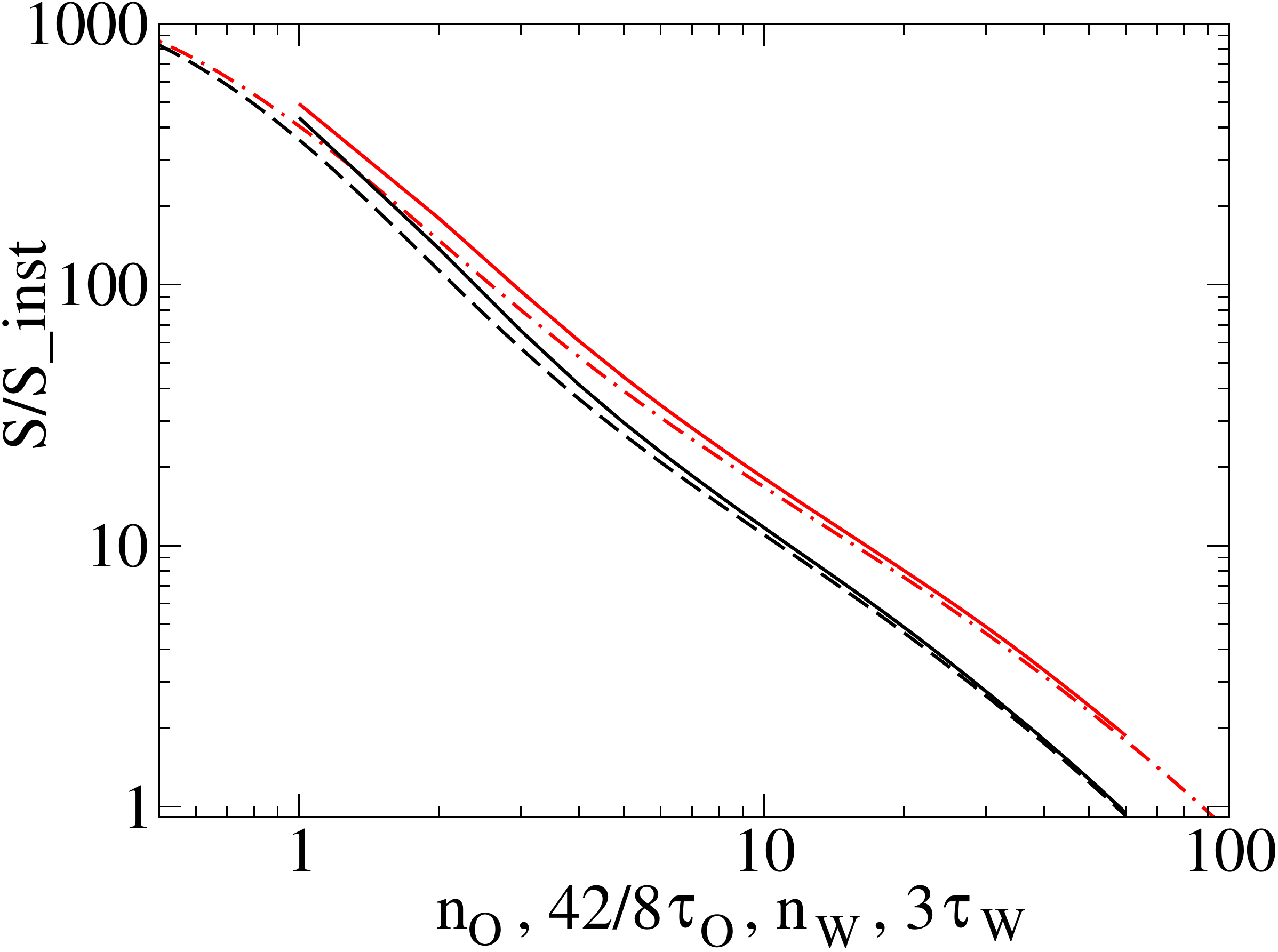}%
\vspace{1cm}
\caption{The evolution of action (\ref{gluonic_S}) (in instanton units) with Wilson 
cooling step $n_W$ (black line) and with Wilson flow time $3\tau_W$ (black 
dashed line) shown on the abscissa, and the variation of the same action (\ref{gluonic_S}) 
(in the same units) with overimproved cooling step $n_O$ (red line) and with
overimproved flow time $(42/8)\tau_O$ shown on the abscissa (red dash-dotted 
line).}
\label{gradient}

\end{figure}

It is worth noting that standard gradient flow or Wilson flow
\cite{Luscher:2009eq,Luscher:2010iy,Luscher:2011bx} ($\epsilon=1$)
can be mapped one-to-one \cite{Bonati:2014tqa} to standard Wilson 
cooling. From Fig.(\ref{gradient}) it is seen that the
same  holds  for over-improved gradient flow 
in the sense that it nicely follows over-improved cooling.

%-----------------------------------------------------------------------------

%% VB this subsection was changed
% the subsection title is identical to one in our previous paper
% for that reason I suggest to change it
%\subsection{The overlap Dirac operator, the NZM band and the UV filtered
%topological density}
\subsection{UV filtered fermionic topological charge density }
\label{sec:overlap}

We consider the near-zero-mode (NZM) band of eigenmodes of the massless overlap 
operator $D$. We use the overlap Dirac operator $D$ 
of the form ~\cite{Neuberger:1997fp,Neuberger:1998wv}
%the Wilson-Dirac operator
%$D_W$ -- is the following zero-mass overlap Dirac operator
\begin{eqnarray}
D(m=0)&=&\frac{\rho}{a}\,\left( 1 + \frac{D_W}{\sqrt{D_W^{\dagger}\,D_W}}
\right) \nonumber \\
&=&\frac{\rho}{a}\,\left( 1 + {\rm sgn}(D_W) \right) \,,
\label{eq:OverlapDirac}
\end{eqnarray}
where $D_W = M - {\rho}/{a}$, $M$ is the hopping term of the
Wilson-Dirac operator and ${\rho}/{a}$ is a negative mass term usually
determined by optimization.
The index of $D$, can be identified 
with the integer-valued topological charge
$Q_{\rm over}$~\cite{Hasenfratz:1998ri}. The non-zero modes 
have vanishing chirality and appear in pairs with modes in the pair
related by $\psi_{\lambda}= \gamma_5 \psi_{-\lambda}$.

For the configurations with $L_\tau=8$ and spatial size $L_s=16$
at $T=T_\chi$ (ensemble I of 50 configurations) we found from the
average square of the number of zero  modes (equal to $\langle Q^2 \rangle$)
$\chi_{\rm top} = (0.71 \pm 0.17) T_\chi^4$.
For the ensemble II of 50 configurations with $L_\tau=8$ and spatial size
$L_s=24$ at $T=1.06 T_\chi$ we found
$\chi_{\rm top} = (0.24 \pm 0.05) T_\chi^4$.
The comparison of these numbers with those of overimproved cooling reported in section II
(the agreement is within 4 or 6 \% for the fourth root of the susceptibility)
specifies the systematical error induced by the cooling method.

The fermionic topological charge density with maximal resolution (down to the lattice
spacing $a$) is defined in terms of the overlap Dirac operator (\ref{eq:OverlapDirac}) as
follows
\begin{equation}
q(x) = - {\rm tr} \left[ \gamma_5 \left( 1
       - \frac{a}{2}\,D(m=0;x,x) \right)\, \right] \,.
\label{eq:TopDensI}
\end{equation}
Using the spectral representation of (\ref{eq:TopDensI}) after diagonalization 
in terms of the eigenmodes $\psi_{\lambda}(x)$ an UV filtered form of the 
density can be defined as a  sum over  
narrow band of NZM 
%near-zero eigenmodes:
\begin{equation}
q_{\lambda_{\rm sm}}(x) = - \sum_{|\lambda| < \lambda_{\rm sm}}
\left( 1 - \frac{\lambda}{2} \right)
\,\sum_c \left( \psi_{\lambda}^c(x)\, ,\gamma_5\, \psi_{\lambda}^c(x) \right)
\label{eq:TopDensII}
\end{equation}
with $\lambda_{\rm sm}$ acting as an UV cutoff. 

The diagonalization of the overlap operator is achieved using a variant 
of the Arnoldi algorithm~\cite{Neff:2001zr}. We had at our disposal between 
20 and 30 non-zero eigenmodes.

While the physical fermion sea is described by the Dirac operator 
implemented with 
antiperiodic temporal boundary conditions, for the purpose of analyzing 
the topological structure it is useful to diagonalize the Dirac operator
subject to continuously modified temporal boundary conditions characterized
by an angle $\phi$,
\beq
\psi(\vec{x},x_4+\beta) = \exp(i\phi)\psi(\vec{x},x_4) \; .
\label{eq:bc1}
\eeq
We have chosen three angles including the case of antiperiodic boundary
condition,
\beq
\phi = \left\{
\begin{array}{ll}
 \phi_1  \equiv -\pi/3\, \\
 \phi_2  \equiv +\pi/3\, \\

 \phi_3  \equiv ~~~~\pi\,
\end{array}
\right\}
\label{eq:bc2}
\eeq
\noindent
ensuring for a single caloron solution that the corresponding fermion zero 
modes become maximally localized at one, but each time at a different one of its 
three constituent dyons. Note that $\phi_3$ corresponds to the antiperiodic
boundary condition.

The construction of the UV smoothed topological charge density
in terms of the eigenvalues and eigenmodes
should be specifically done for the three boundary conditions:
\beq
q_{i,\lambda_{\rm sm}}(x) = - \sum_{|\lambda| < \lambda_{\rm sm}}
\left( 1 - \frac{\lambda}{2} \right)
\,\sum_c \left( \psi_{i,\lambda}^c(x)\, ,\gamma_5\, \psi_{i,\lambda}^c(x) \right) \, ,
\label{eq:truncated_density}
\eeq
where $i =1,2,3$ enumerates the three boundary conditions defined by 
\Eq{eq:bc2}.

For the configurations with $L_\tau=8$ and spatial sizes $L_s=16$ 
(the ensemble I of 50 configurations at $T=T_\chi$) we take 
$\lambda_{\rm sm}= 331$ MeV. 
This is the minimal spread of $|\lambda|$ among the 20 non-zero eigenvalues 
we have found per configuration (minimal with respect to all 50 configurations 
and 3 boundary conditions). 
For the configurations with $L_\tau=8$ and spatial sizes $L_s=24$ 
(the ensemble II of 50 configurations at  $T=1.06 T_\chi$) we take 
$\lambda_{\rm sm}= 254$ MeV which is the minimal spread of $|\lambda|$ among
30 eigenvalues per configuration which we have determined by diagonalization.
The actual number of non-zero modes falling into this interval and included 
into the definition \Eq{eq:TopDensII} fluctuates from configuration to 
configuration because in the present analysis an eigenvalue cut-off is 
applied instead of a fixed number of modes.

We remark that in the case of ensemble I we have, besides of the cutoff 
$\lambda_{\rm sm}=331$ MeV, also considered the smaller cutoff 
$\lambda_{\rm sm}=254$ MeV known from ensemble II.
This implies that the construction of the topological density including 
less non-zero modes than for the larger cutoff. This will allow us to 
discuss the effect of changing the cutoff for ensemble I (representing
the lower temperature) and to compare between the two temperatures 
(applying the smaller cutoff both to ensemble I and ensemble II).

The localization of the topological charge within a given charge density
filling a lattice configuration can be measured by the inverse participation
ratio ${\mathrm{IPR}}$ which varies between the extremes 1 (totally 
delocalized) and $V_4$ (fully localized). It is defined as
\beq
{\mathrm{IPR}} = V_4 \frac{\sum_x |q(x)|^2}{(\sum_x |q(x)|)^2}~~.
\label{eq:localization}
\eeq
$V_4$ is the four dimensional volume.
Any subvolume $fV_4$ equally filled results in ${\mathrm{IPR}}=1/f$.
We have analyzed the IPR \Eq{eq:localization} of the three types of fermionic
toplogical charge density corresponding to the three boundary conditions,
at both temperatures for the same cutoff, and for the lower temperature 
(at $T_\chi$) with two different cutoffs. The result is presented in Table I.

%---------------------------------------------------------------------
\begin{table*}[ht]
\begin{center}
\vspace*{0.5cm}
\begin{tabular}{|l|c|c|c|}
\hline
type of            & ensemble I & ensemble I & ensemble II \\
boundary condition & $\lambda_{\rm sm} = 331$ MeV & 
                     $\lambda_{\rm sm} = 254$ MeV &
                     $\lambda_{\rm sm} = 254$ MeV \\
\hline
1-st type b.c. & $2.279(30)$ & $2.373(36)$ & $3.312(76)$ \\
\hline
2-nd type b.c. & $2.282(29)$ & $2.428(36)$ & $3.259(84)$ \\
\hline
3-rd type b.c. & $2.354(35)$ & $2.493(45)$ & $8.436(964)$ \\ 
\hline
\end{tabular}
\vspace*{0.5cm}
\caption{Inverse Participation Ratio (IPR) of the fermionic topological 
charge density at the selected fermionic cutoffs $\lambda_{\rm sm}$, for 
three types of boundary conditions.}
\end{center}
\end{table*}
\label{tab:tableI}
%----------------------------------------------------------------------------

We can conclude, that at equal cutoff $\lambda_{\rm sm}=254$ MeV
for antiperiodic boundary conditions the localization is 3-times
bigger at higher temperature $T=1.06 T_\chi$ compared to $T = T_\chi$.
In addition, at the higher temperature, antiperiodic boundary conditions 
result in a localization by a factor 2.5 stronger than the other two 
boundary conditions.

At the lower temperature $T_\chi$, the lower cutoff $\lambda_{\rm sm}=254$ MeV 
leads only to a minor increase of localization by a few percent than
the larger cutoff $\lambda_{\rm sm}=331$ MeV. The effect of changing 
boundary conditions is at a similar level.

\begin{figure*}[htb]
\centering 
\includegraphics[width=0.41\textwidth]{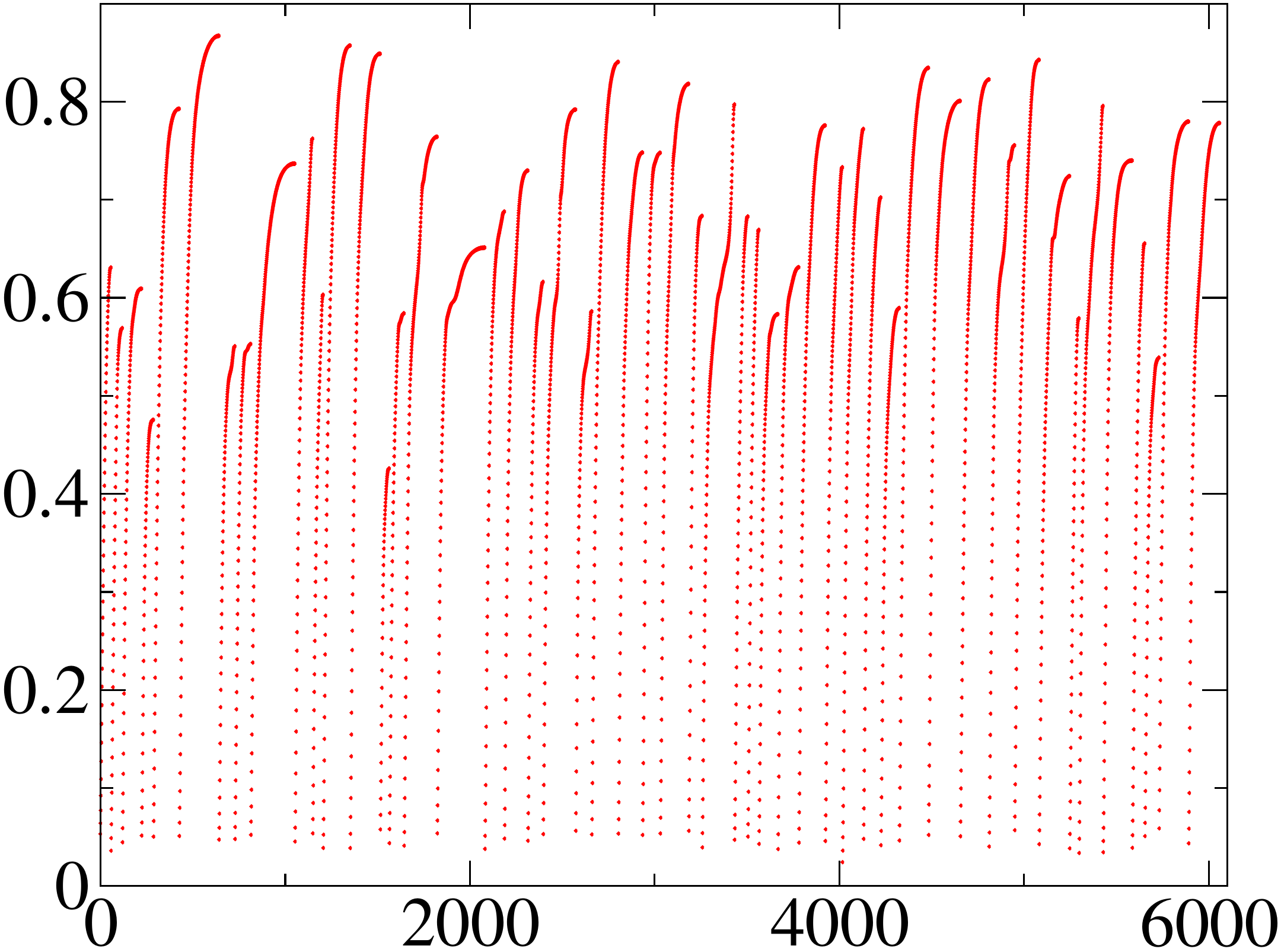}%
\hspace{0.5cm}
\includegraphics[width=0.41\textwidth]{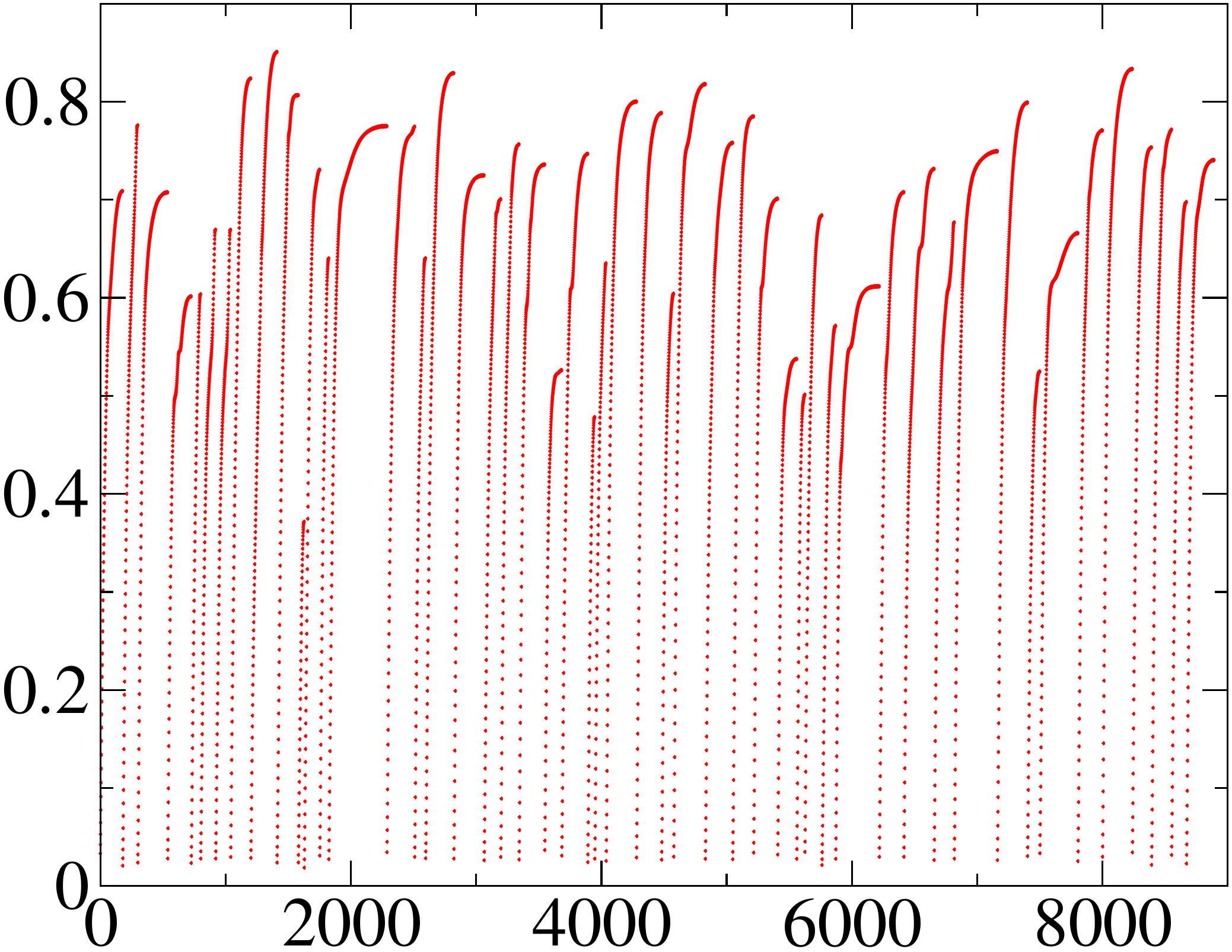}
\caption{The evolution of the cosine $\cos(\theta)$, the cosine of an ``angle''
between gluonic and fermionic topological charge density (see text),
shown as function of  common flow time ($\tau_O/\Delta \tau$), consecutively for all 50 configurations
of the ensemble I (left panel) and of the ensemble II (right panel).}
\label{cos}
\end{figure*}

We are now going to compare the fermionic topological density $q_f(x)$, truncated 
according to (\ref{eq:truncated_density}) with a fixed cut-off $\lambda_{\rm sm}$
and {\it averaged} over the boundary conditions, with the gluonic topological 
density $q_g(x)$ (\ref{eq:qdensity}), provided a suitable amount of over-improved 
gradient flow has been applied to the gauge field configuration. 
Before we can compare, both topological densities 
should be corrected (by a shift~\cite{Bruckmann:2006wf}) to have a vanishing volume 
average,
\beq
q_g(x) \to q_g(x) - \bar{q}_g 
\eeq
and 
\beq
q_f(x) \to q_f(x) - \bar{q}_f \; .
\eeq
The optimal matching between the two (corrected) topological densities 
is achieved when the properly normalized ``scalar product'' between the two 
topological densities $q_g-\bar{q}_g$ and $q_f-\bar{q}_f$, expressed by the
cosine between ``direction vectors'',
\beq
\cos(\theta) = \frac{((q_g-\bar{q}_g),(q_f-\bar{q}_f))}
{|q_g-\bar{q}_g| |q_f-\bar{q}_f|}
\label{eq:cosine}
\eeq
passes a maximum. Fig.(\ref{cos}) shows the rise of the cosine towards the
maximum for the 50 individual configurations of both ensembles, ensemble I 
(left panel) and ensemble II (right panel), as function of the flow time 
$\tau_O$, until the flow is stopped at the maximum of the cosine.

\begin{figure*}[htb]
\centering
\includegraphics[width=0.41\textwidth]{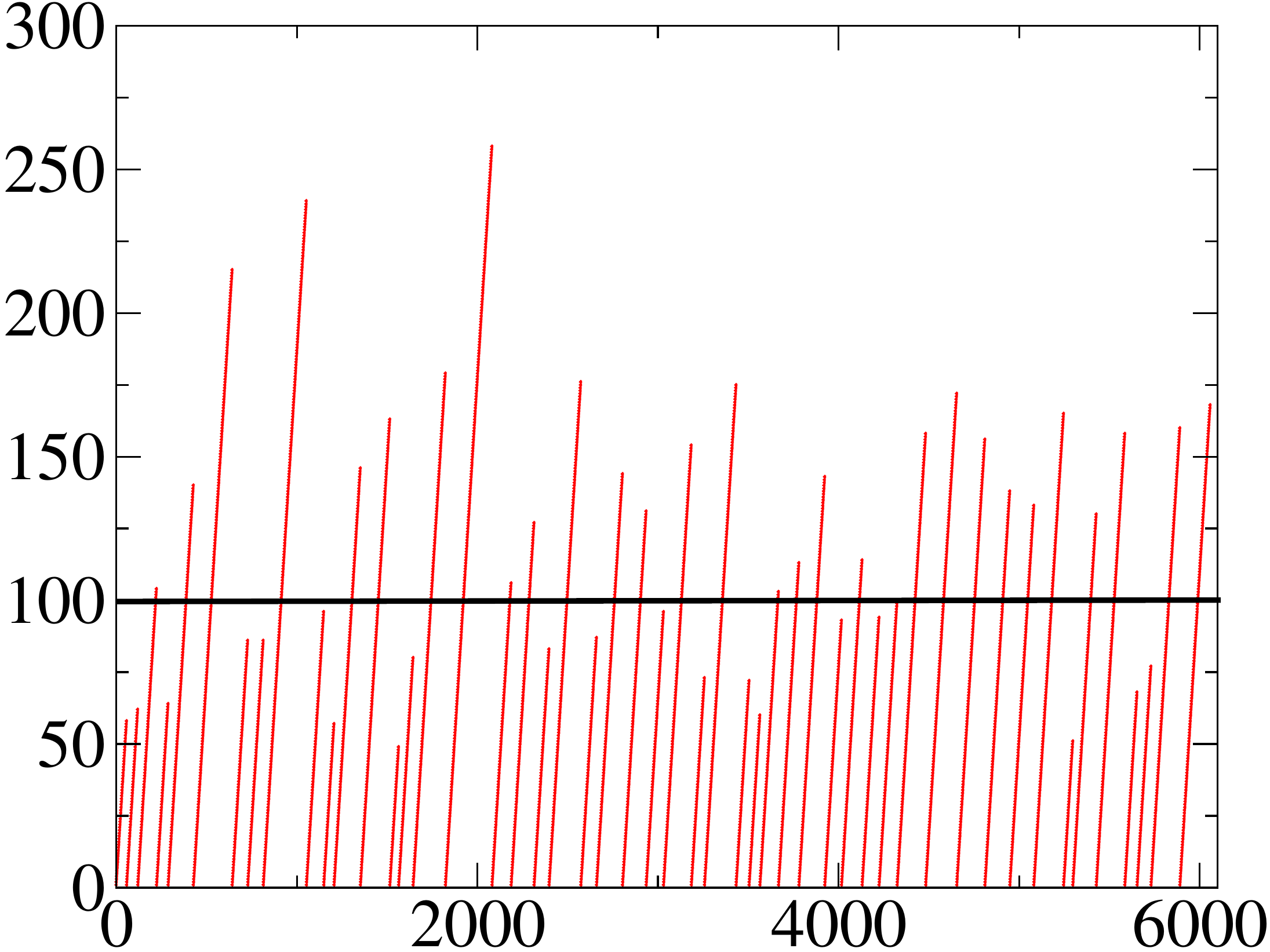}%
\hspace{0.5cm} 
\includegraphics[width=0.41\textwidth]{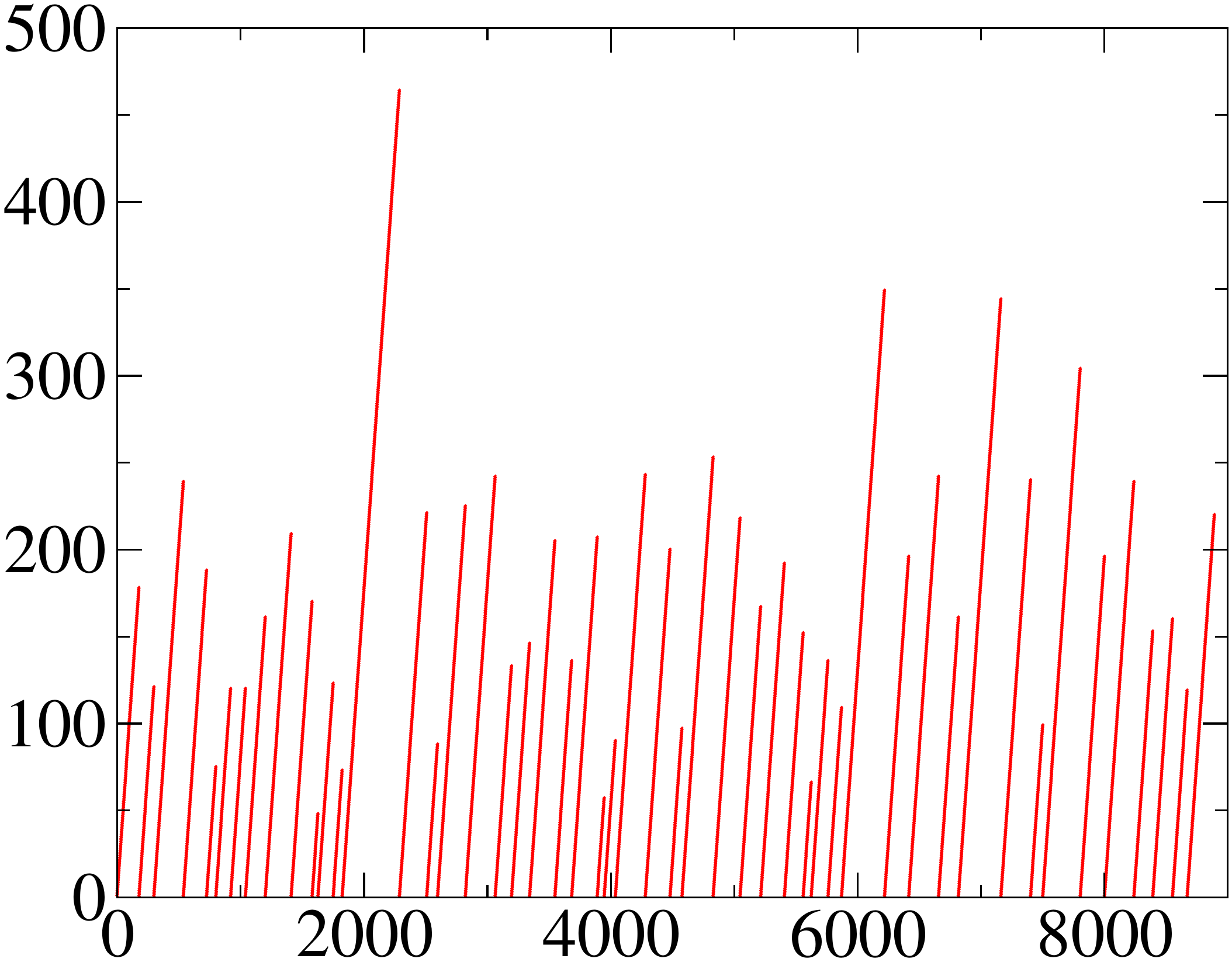}
\caption{The number of steps of over-improved gradient flow 
($\tau_O/\Delta \tau$) (with step size $\Delta \tau =0.02$) growing
until the cosine reaches the maximum, i. e. the best matching is achieved
between the gluonic topological charge density and the fermionic topological 
charge density (averaged over boundary conditions).
Shown for all 50 configurations of the ensemble I (left panel)  
and ensemble II (right panel).}
\label{lengths}
\end{figure*}

Fig.(\ref{lengths}) shows the step number (corresponding to a finite flow 
time step $\Delta \tau = 0.02$ adopted) linearly growing until the 
over-improved gradient flow is stopped. This necessary step number fluctuates 
from configuration to configuration. Let us note that the average step number 
($\approx 100$) of the over-improved gradient flow corresponds to a flow time 
$\tau_O \approx 2$. This is the time required to let the diffusion 
length~\footnote{diffusion length given in lattice units $a$} 
of gradient flow, $\sqrt{8 \tau_O}$, grow to the size of a dyon  
$N_\tau/(2\pi \nu_d)$ where $\nu_d=\nu_1=\nu_2=\nu_3=1/3$. This would 
adequately describe the fractional charges in the case of maximally 
nontrivial holonomy (vanishing average Polyakov loop). This average step 
number is shown by a horizontal line in the left panel of Fig.(\ref{lengths}) 
which represents the smoothing of the ensemble I.
The averaged value of dimensionless quantity 
$t_O^2 \cdot \frac{1}{2} \Tr (F_{\mu\nu} (x)~F_{\mu\nu} (x))$ (with the
dimensionful flow time $t_O = a^2 \tau_O$) is equal to 0.37 (in the case 
of ensemble I) and 0.34 (for the ensemble II). This fits remarkably well
to the mostly used stopping criterion for Wilson flow.  

\begin{figure*}[htb]
\centering
\includegraphics[width=0.41\textwidth]{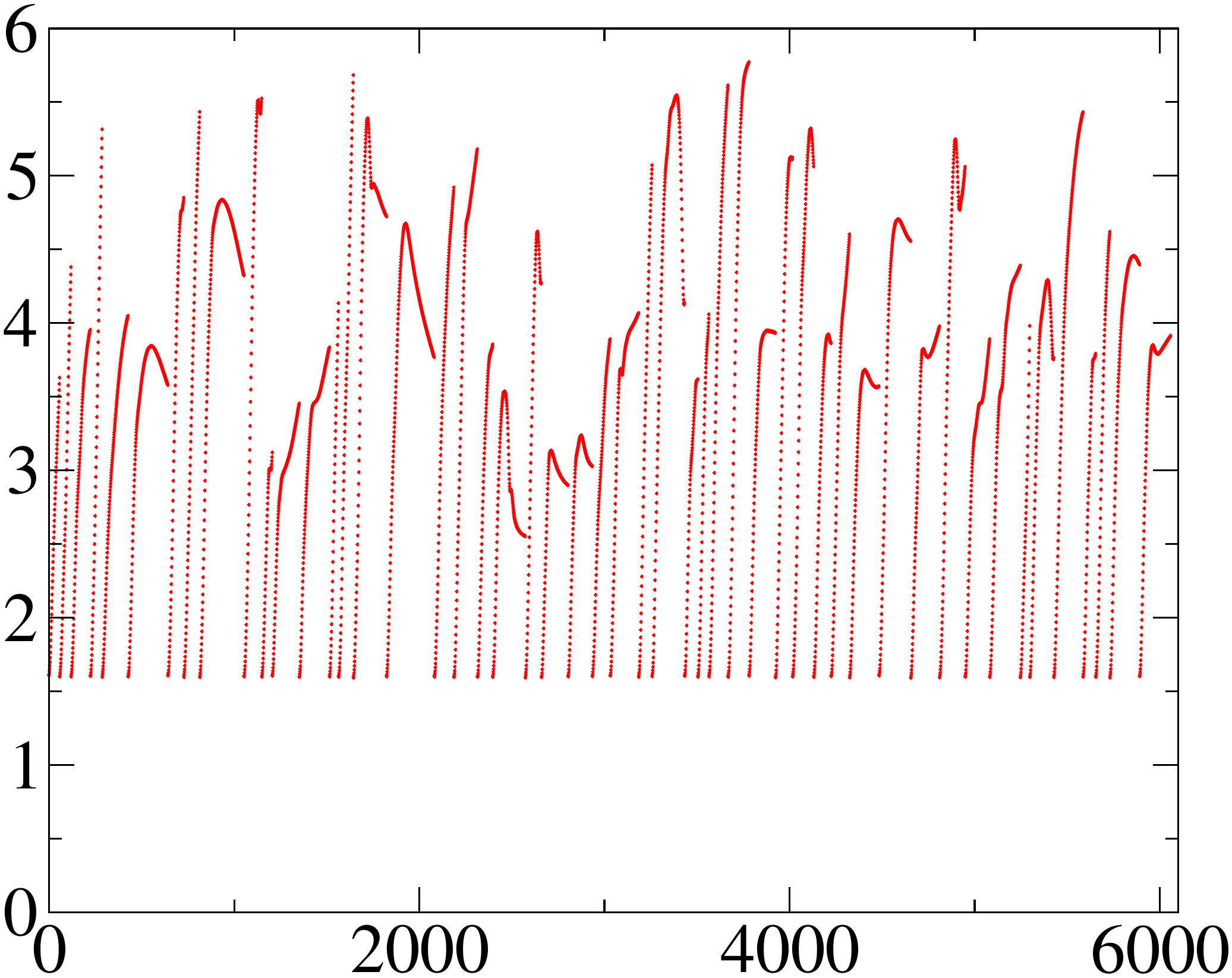}%
\hspace{0.5cm} 
\includegraphics[width=0.41\textwidth]{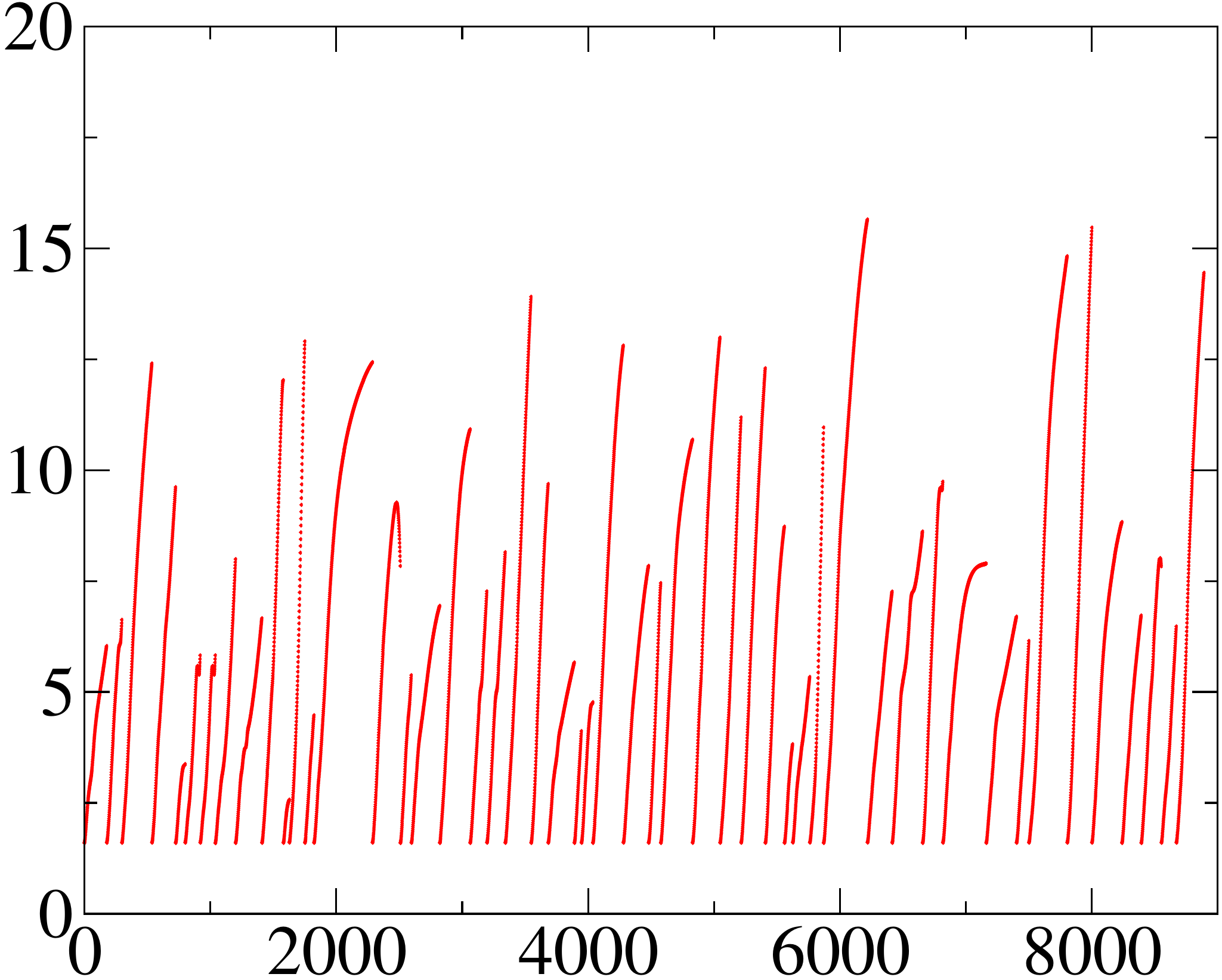}
\caption{The  inverse participation ratio ${\mathrm{IPR}}$ evolving in the 
course of over-improved gradient flow ($\tau_O/\Delta \tau$) until the 
best matching between the gluonic topological charge density and the 
fermionic topological charge density (averaged over boundary conditions) 
is achieved. Shown for all 50 configurations of the ensemble I (left panel)  
and ensemble II (right panel).}
\label{IPR}
\end{figure*}

Fig.(\ref{IPR}) shows for each configuration how the inverse participation
ratio ${\mathrm{IPR}}$ evolves in the course of gradient flow, until the 
average fermionic density is optimally matched. The filling fraction $f$ 
changes from $1/3$ to $1/10$ with the increase of temperature (see also
the IPR for the fermionic topological charge density). 

Topologically non-trivial clusters filtered out with the three truncated 
fermionic densities (\ref{eq:truncated_density}) (each corresponding to 
one type of fermionic boundary condition) will be separately considered 
in the following as topologically different objects (dyon candidates).
Their localization was already described by a corresponding ${\mathrm{IPR}}$
(see Table I).
They may appear either isolated or forming compounds with other dyons 
(dyon-dyon pairs) or antidyons (dyon-antidyon pairs), including the 
possibility to recombine into calorons.

%-----------------------------------------------------------------------------

%----------------------------------------------------------------------
 \section{Results of cluster analysis for the QCD ensembles}
%----------------------------------------------------------------------
\label{sec:ensemble_results}

In the following we will analyze our two QCD ensembles of thermalized 
configurations along the lines sketched above. In order to identify 
topological clusters of the lattice gauge fields with the help of the 
low-lying spectrum of the overlap operator, we used a fixed cut on 
eigenvalues $\lambda_{\rm sm}= 331$ MeV for configurations with 
$L_s=16$, $T=T_\chi$ (see above). For configurations with $L_s=24$ 
(50 configurations at $T=1.06T_\chi$) we take $\lambda_{\rm sm}= 254$ MeV.
In all three sectors (selected by the angles $\phi_i$) the fermionic 
topological density is constructed.

For the purpose of detecting gluonic features of (anti)dyon excitations
among such clusters we have made the configurations undergo the procedure
of over-improved gradient flow until the gluonic topological density 
profiles optimally matched the (averaged over sectors) fermionic one, 
analogously to what we did in our previous paper \cite{Ilgenfritz:2013oda} 
where we followed the concept of an equivalent filtering as developed in
\cite{Bruckmann:2006wf,Ilgenfritz:2008ia,Bruckmann:2009vb}. This filtering 
in particular acting (smoothing) the local holonomy, while the gluonic 
topological density fits to the sector-averaged fermionic density, anyway, 
by construction.

First we have applied the same cluster analysis as in our previous paper 
\cite{Ilgenfritz:2013oda} with a variable lower cutoff $q_{\rm cut}>0$ 
in order to characterize the cluster properties of the three density 
functions \Eq{eq:truncated_density} in the thermal ensembles.   

Let us summarize here the idea of the cluster algorithm. In a first step - 
for each of the three fermionic boundary conditions \Eq{eq:bc2} - the 
algorithm identifies the lattice points forming the interior of {\it all} 
clusters (the so-called ``topological cluster matter'') defined by the 
condition $|q(x)| > q_{\rm cut}$. The crucial second step is to enquire 
the connectedness between the lattice points in order to form individual 
extended clusters out of this ``cluster matter''.
Neighbouring points with $|q(x)|$ above threshold and sharing the same sign 
of the topological charge density are declared to belong to the same cluster.
The cutoff $q_{\rm cut}$ has been chosen such as to resolve the given
``continuous'' distribution $q(x)$ into a {\it maximal number} of internally 
connected and mutually separated clusters. The cutoff value has been 
independently adapted for each configuration.
 
For ensemble I, the difference between the two cut-off values $\lambda_{\rm sm}$
is small.
Between the two levels of gradient flow, the resulting number of all
clusters $N_{\rm cl}$ coincide within erors.
On the other hand, the same cut-off $\lambda_{\rm sm}= 254$ MeV applied
to lattices with different volumes 
($V_4(II)/V_4(I) = (3/2)^3/(1.06)^4\approx 2.67$)
gives us the possibility to understand the size of finite volume effects: 
all extensive quantities (as it can be seen from the Table I) are differing
from each other by a factor of order 2.

There are two conditions to check the {\it dyonic nature} of isolated 
clusters, namely (1) an integrated topological charge close to $\pm$ 
one third {\it and} (2) the existence of a point inside where two 
eigenvalues of the local holonomy become degenerate. 

\begin{figure*}[htb]
\centering 
\vspace{2.5cm}
\includegraphics[width=0.41\textwidth]{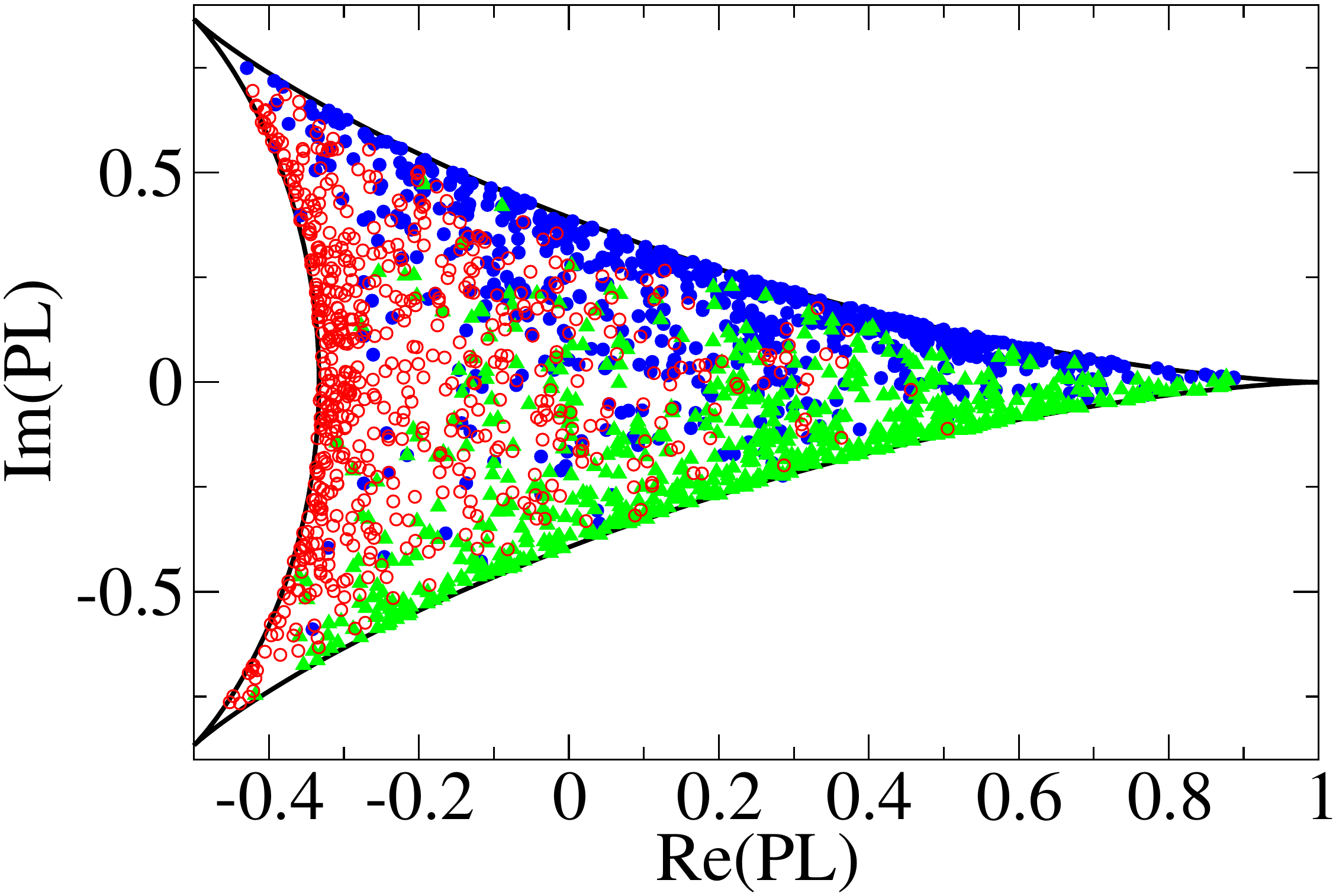}
\hspace{0.5cm} 
\includegraphics[width=0.41\textwidth]{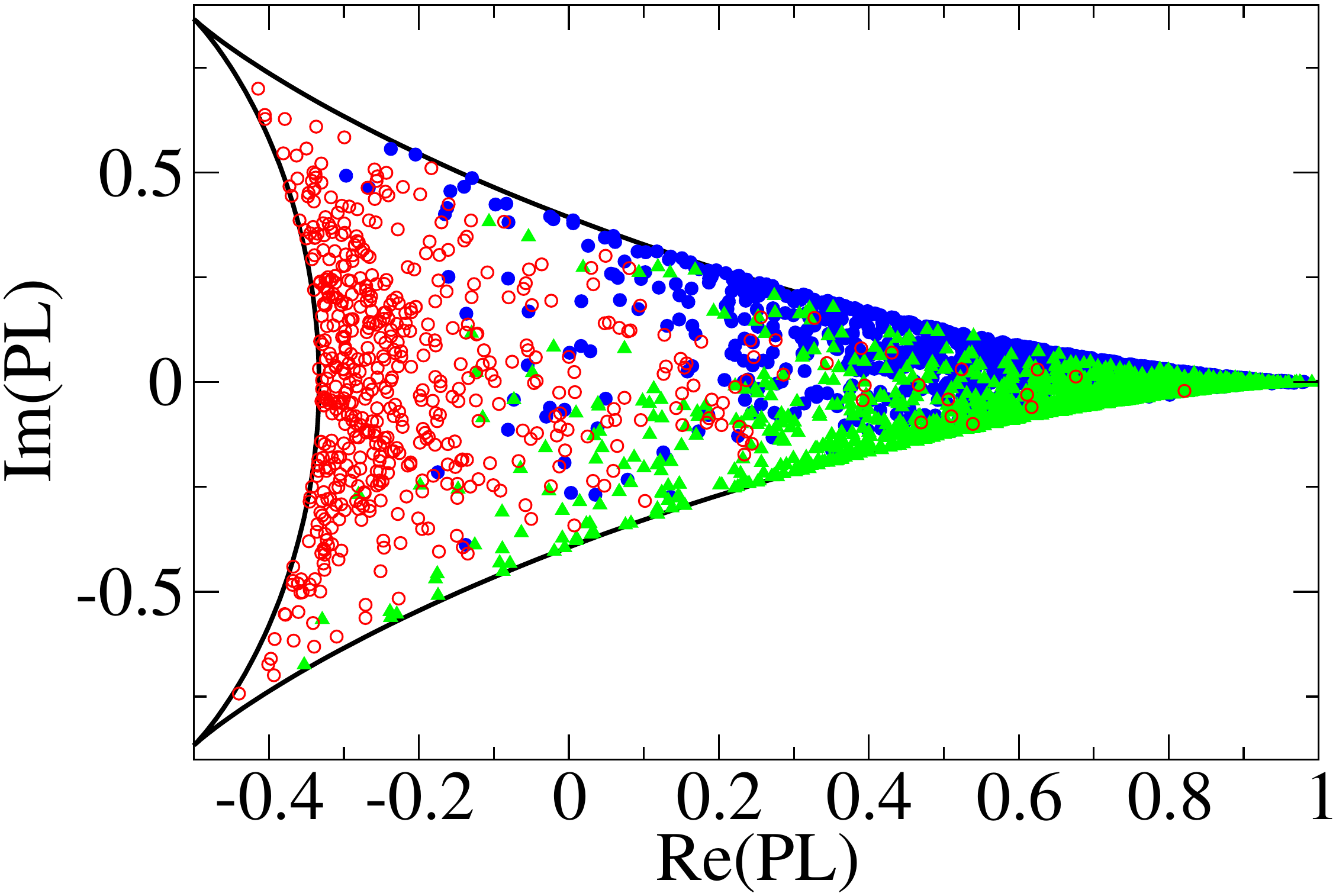}
\vspace{0.5cm}
\includegraphics[width=0.41\textwidth]{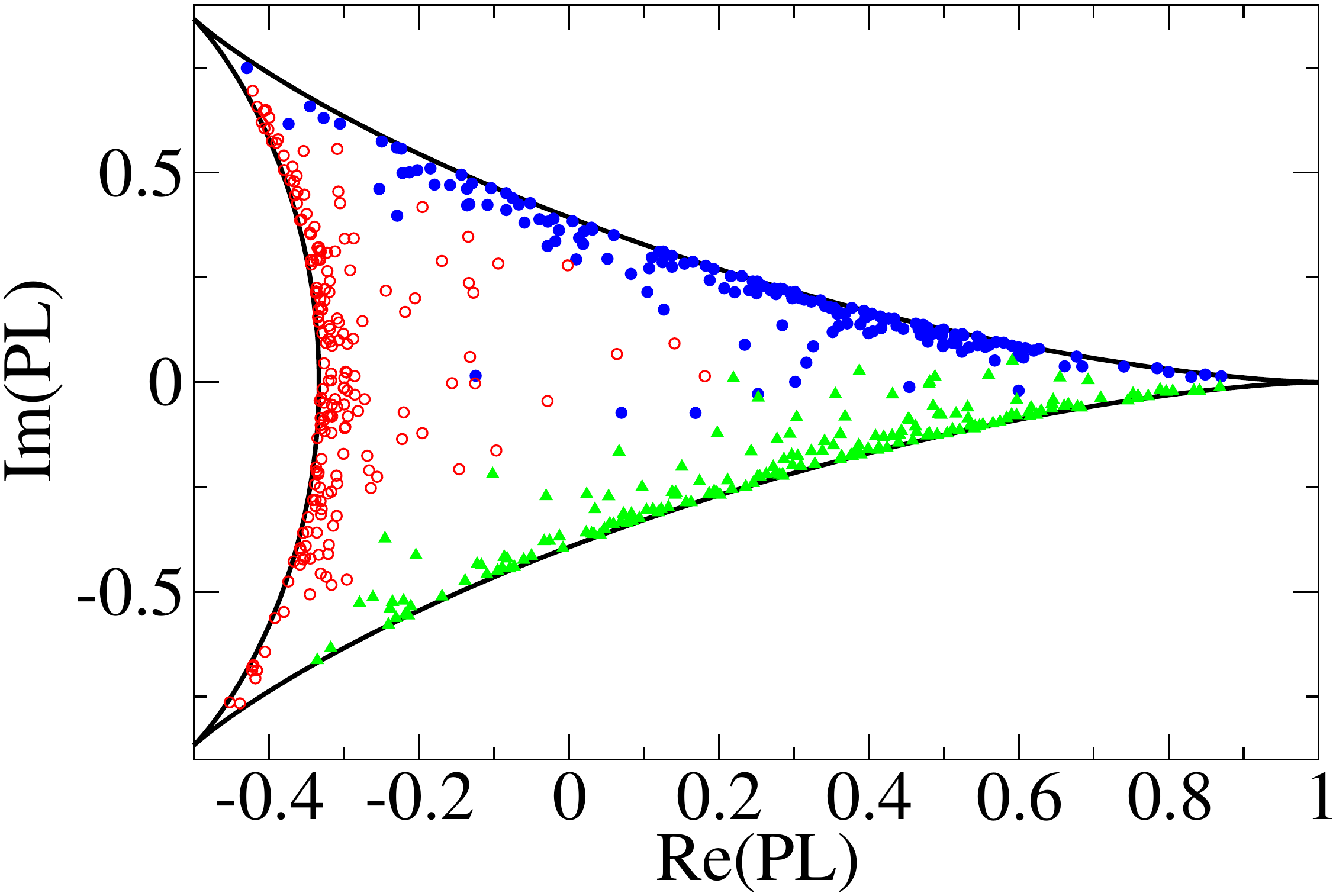}
\hspace{0.5cm} 
\includegraphics[width=0.41\textwidth]{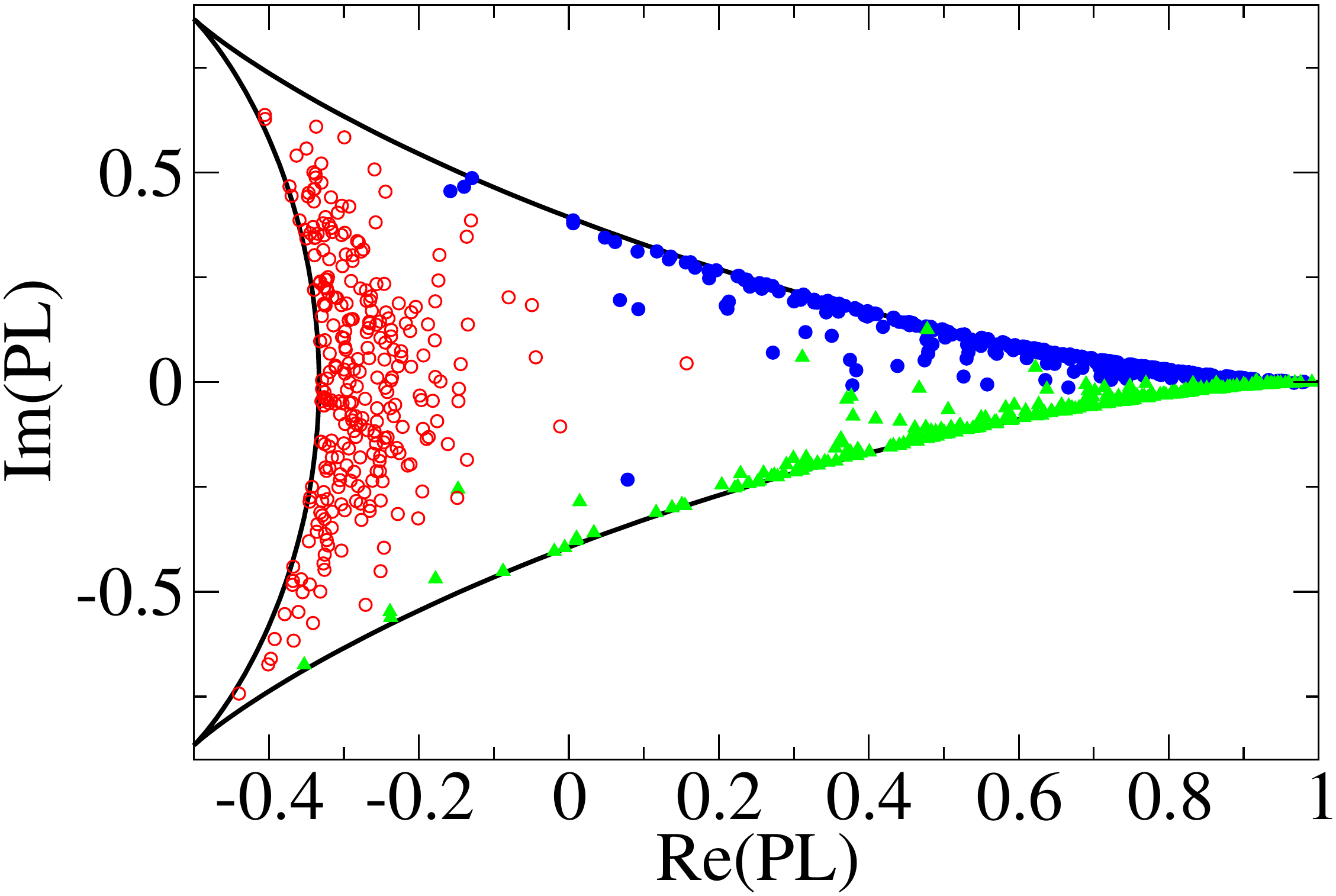}\\
\caption{Scatter plots of Polyakov loop $PL$ (for configurations having
undergone the over-improved gradient flow procedure)
in in the centers of all clusters (upper row) and  of clusters selected 
to contain monopoles(lower row). The clusters are separated according to 
the type of boundary condition for the overlap near-zero modes.
For clusters of first type the Polyakov loop is shown by green triangles,
for clusters of second type - by blue filled circles,
for clusters of third type - by red open circles,
This scatter plots show all clusters of all 50 configurations of size  
$L_s=16$ and $T=T_\chi$ on the left (ensemble I) and size $L_s=24$ and 
$T=1.06 T_\chi$) on the right (ensemble II).}
\label{plincl}
\end{figure*}

If the holonomy is represented by the local 
Polyakov loop, this second condition corresponds to a position 
of one point in the cluster's Weyl plot close to one of its three sides. 
According to the type of fermionic boundary condition applied, one can 
anticipate which side of the Weyl plot should be approached: the left side 
corresponding to antiperiodic boundary conditions $\phi_3$, the upper right 
side corresponding to $\phi_2$, the lower right side corresponding to $\phi_1$.
Among all points belonging to a cluster we have searched for a minimum of 
the distance to the corresponding side of the Weyl plot. In this way we have 
defined which point should be considered as the ``center of the cluster'' 
(compare Fig.(\ref{plincl}). 

In order to find the integrated topological charge of the clusters we use 
the gluonic topological charge density as it has appeared after applying
the gradient flow process required to match the gluonic topological charge 
density to the fermionic one.
In this case we find clusters of the gluonic topological charge density 
in the same way as we found before clusters of fermionic topological charge 
density according to three different types of fermionic boundary conditions. 

If for example the extrema of the clusters of fermionic topological charge 
density for the first type of b.c. (with time-like monopole links inside) 
fall into a cluster of gluonic topological charge, we say that this gluonic 
topological cluster contains a dyon of first type. The same procedure is 
applied to identify the character as being another type of dyons. 
This gives us the possibility to sort all gluonic topological clusters into
the following categories: as full calorons (three different dyons inside), 
as 3 types of dyon pairs (two different dyons inside), as 3 types of dyons 
(one dyon inside), or as background (no dyons inside). 

\begin{figure}[htb]
\centering 
\includegraphics[width=0.41\textwidth]{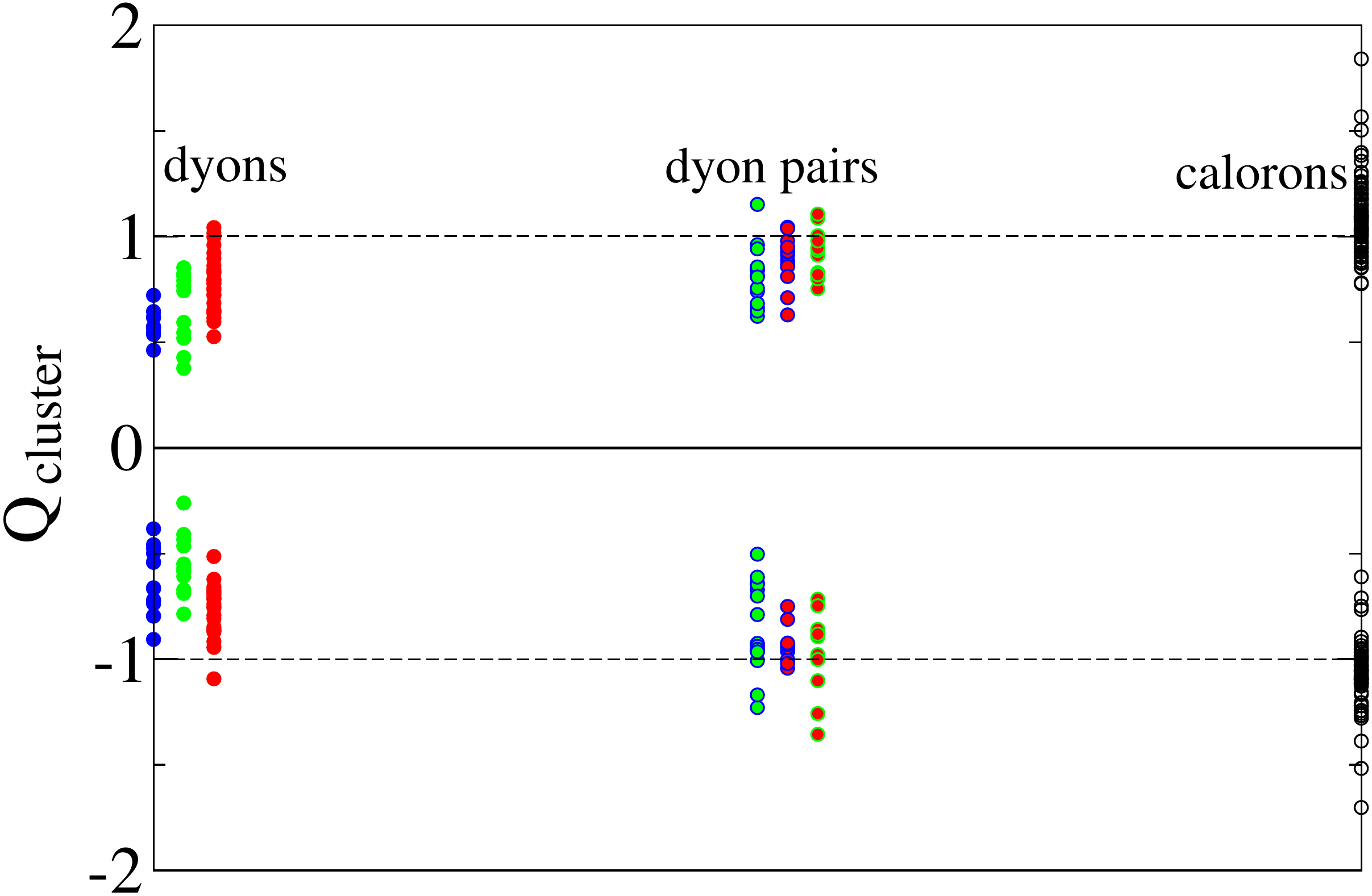}%
\vspace{1cm}
\caption{The integrated topological charges of gluonic topological clusters
interpreted as  3 types of dyons,  3 types of dyon pairs and full calorons 
(analyzed for the ensemble II). The color code refers to the type of dyons.}
\label{topcharge}
\end{figure}

For isolated dyons and dyon pairs and for full calorons we find the integrated 
topological charges the same way as for isolated dyons and calorons in the 
case of SU(2) objects \cite{Ilgenfritz:2004zz}.
The results can be seen on Fig.(\ref{topcharge}) for the ensemble II.

There is another question to be checked: correlation with the time-like
currents of Abelian monopoles in the Abelian projection after MAG. 
This is particularly interesting at much higher temperature where Abelian 
monopoles become thermal (cyclic) and almost static (exclusively timelike).
Even close to the transition like in our case this might {\it sharpen} the 
test for clusters being really dyons or antidyons.

All quantitative data on the correlations between topological charge 
density and MAG monopole content is presented in Table II.

%---------------------------------------------------------------------
\begin{table*}[ht]
\begin{center}
\vspace*{0.5cm}

Clusters obtained with lowest overlap modes for $16^3\times 8$ configurations
\begin{tabular}{lll|c|c|c|c|c|c|}
\hline
 Type of clusters & $V_{cl}$ & $V_{clmon}$ & $ N_{cl} $ & $N_{clmon}$ & $ N_{mon} $ & $N_{moncl}$
  & $ N_{loop} $ & $N_{loopcl}$ \\
\hline
1-st type  clusters  &$6.7(8)\%$ & $6.1(8)\%$ & $ 12.9(3) $ & $3.5(2)$ & $ - $ & $32(3)$& $ - $ & $5.4(4)$ \\
\hline
2-nd type clusters &$6.0(7)\%$ & $5.2(7)\%$ & $ 13.1(3) $ & $3.7(2)$ & $ - $ & $30(2)$& $ - $ & $5.4(4)$ \\
\hline
3-d type clusters  &$6.1(8)\%$ & $5.5(8)\%$ & $ 12.0(4) $ & $3.6(2)$ & $ - $ & $31(2)$& $ - $ & $5.1(4)$ \\
\hline
\hline
All clusters ($\lambda_{\rm sm}= 331$ MeV)&$11(1) \%$ & $10(1) \%$ & $ 38 (1) $ & $10.8 (6)$ &
$ 129 (4) $ & $45 (3)$& $ 8.7 (5) $ & $6.4 (4)$ \\
 All clusters  ($\lambda_{\rm sm}= 254 $ MeV) &$ 12(1)\%$ & $ 11(1)\%$ & $  32(1) $ & $ 9.6(6)$ &
$  120(4) $ & $ 42(3)$& $  8.1(5) $ & $ 5.8(4)$ \\
 \hline
\end{tabular}
\vspace*{0.5cm}

Clusters obtained with lowest overlap modes for $24^3\times 8$ configurations
\begin{tabular}{lll|c|c|c|c|c|c|}
\hline
 Type of clusters & $V_{cl}$ & $V_{clmon}$ & $ N_{cl} $ & $N_{clmon}$ & $ N_{mon} $ & $N_{moncl}$
  & $ N_{loop} $ & $N_{loopcl}$ \\
\hline
1-st type  clusters  &$5.1(3)\%$ & $4.4(4)\%$ & $ 24(1) $ & $5.8(3)$ & $ - $ & $59(3)$& $ - $ & $9.9(6)$ \\
\hline
2-nd type clusters &$6.2(6)\%$ & $5.5(6)\%$ & $ 24(1) $ & $5.5(3)$ & $ - $ & $64(4)$& $ - $ & $10.6(7)$ \\
\hline
3-d type clusters  &$1.8(2)\%$ & $1.5(2)\%$ & $ 11(1) $ & $5.7(4)$ & $ - $ & $40(3)$& $ - $ & $6.7(6)$ \\
\hline
\hline
All clusters  ($\lambda_{\rm sm}= 254 $ MeV)&$8.0(6)\%$ & $7.0(6)\%$ & $ 59(2) $ & $17.0(8)$ &
$ 193(8) $ & $76(5)$& $ 20.2(8) $ & $12.1(7)$ \\
\hline
\end{tabular}
\label{tabdata}
\vspace*{0.5cm}
\caption{Results of the cluster analysis using low-lying overlap operator
modes with three kinds of boundary conditions. All numbers indicate averages 
per configuration. The pure statistical errors are given in parentheses.
We denote with
$V_{\rm cl}$      - the volume fraction occupied by all topological clusters,
$V_{\rm cl~mon}$  - the volume fraction occupied by clusters identified to
                    contain time-like magnetic monopoles,
$N_{\rm cl}$      - the number of all clusters per configuration,
$N_{\rm cl~mon}$  - the number of clusters identified to contain time-like
                    magnetic monopoles,
$N_{\rm mon}$     - the overall number of dual timelike links carrying 
                    monopole currents,
$N_{\rm mon~cl}$  - the number of dual timelike links carrying monopole 
                    currents found inside topological clusters,
$N_{\rm loop}$    - the overall number of thermally closed monopole worldlines,
$N_{\rm loop~cl}$ - the number of thermally closed monopole worldlines 
                    piercing topological clusters.
The first part of the table is related to the ensemble I 
($L_s=16$ at $T=T_\chi$).
Here we show the effect of changing the cut-off from $\lambda_{\rm sm}= 331$ 
MeV (derived for 20 eigenmodes) to a smaller cut-off $\lambda_{\rm sm}= 254$ 
MeV (implying more iterations of gradient flow).
The second part of the table is related to the ensemble II 
($L_s=24$ at $T=1.06 T_\chi$), where the cut-off $\lambda_{\rm sm}= 254$ MeV 
(derived for 30 eigenmodes) has been applied.} 
\end{center}
\end{table*}
\label{tab:tableII}
%----------------------------------------------------------------------------

Our main results on the correlation of low-lying modes of the overlap Dirac
operator (as represented by the clusters of fermionic topological charge)
with the Abelian monopoles of MAG are as follows. For the ensemble I 
(ensemble II) the topological clusters occupy about 11\% (8\%) of the lattice 
4-volume, whereas topological clusters constrained to contain static MAG 
monopole currents cover 10\% (7\%) of the lattice volume, but the latter 
contain about 35\% (40\%) of the time-like dual links carrying MAG monopoles.
Inside those topological clusters which are pierced by MAG monopoles, the 
density of monopoles is about 5 (9) times larger than outside these clusters. 
These numbers become even more pronounced if one counts not just the time-like 
monopole currents (dual links) in topological clusters  but the numbers of 
thermal monopoles piercing topological clusters. Around 75\% (60\%) of thermal 
(thermally winding) monopoles are seen piercing topological clusters.

We expect that the topological clusters detected with antiperiodic
boundary conditions (in our case with a real-valued average Polyakov loop)
can be viewed as related to more heavy dyons: at the higher temperature
$T= 1.06T_\chi$ (in the ensemble II) they are expected to become statistically 
suppressed because of their higher action in comparison with the other 
constituents of a caloron.
We can estimate this suppression quantitatively by measuring the abundance 
of thermal monopoles piercing topological clusters of third type compared 
to those piercing topological clusters of first or second type.
Such thermal monopoles, if they are correlated with clusters of topological
charge, we are inclined to associate with physical dyons.
We found the proportion $~~9.9 : 10.6 : 6.7~~$ (see the lower subtable).
Thus, the heavier caloron constituent clusters are really suppressed already
when the temperature is exceeding $T_\chi$ by only few percent.

Furthermore, the correlation between the local Polyakov loop on one side 
and the dyon nature of clusters of topological charge on the other is 
increased if the clusters are constrained to be coinciding with Abelian 
monopoles. The Polyakov loop tests the degeneracy of holonomy eigenvalues 
in the cluster centers which are (by our definition above) distinguished 
among all the cluster points by the condition to have a minimal distance 
from the sides of the Weyl plot. 

We show the scatter plot of local Polyakov loops measured in the centers 
of those clusters which are associated with magnetic monopoles.
Since the clusters are labelled (see the color code) by one of the three 
boundary conditions for the fermionic modes (used to define the fermionic 
topological charge density), the scatter plot over the Weyl plot  
\Fig{plincl}~ shows the different regions of population. The tendency 
of the Polyakov loop in the cluster centers to concentrate along the 
corresponding sides of the Weyl plot is much more pronounced than it 
was for all clusters (corresponding picture before).
We remind that the Polyakov loop is measured for the gluonic field that 
has emerged from gradient flow at maximal matching of gluonic and fermionic
topological densities.

\section{Conclusions}
%--------------------
\label{sec:conclusions}
For lattice QCD we have discussed the dyonic signatures of clusters of
topological charge selected by three types of temporal boundary conditions 
applied to the overlap modes used in the fermionic definition of topological 
density. In contrast to our previous paper on $SU(3)$ gluodynamics ~\cite{Bornyakov:2014esa}
here  for the construction 
of fermionic topological densities not the number of pairs of near-zero modes is fixed but the eigenvalue cutoff.
The thermal lattice gauge fields were generated close to the crossover
temperature. 
Topological clusters to be considered as candidates to be dyons
were established by filtering, i.e. restricting to low-lying modes of the 
overlap Dirac operator with specific boundary conditions.  
Additionally, we have applied to the gluonic lattice fields the procedure
of overimproved gradient  flow (instead of overimproved cooling in ~\cite{Bornyakov:2014esa})
after which a similar pattern of clusters 
occurs within the gluonic topological charge distribution
(however averaged over boundary conditions).
We looked for distributions of local Polyakov loop and searched for MAG 
monopole currents in the gradient-flow-smeared gluonic fields.
We found clear correlations of the topological clusters
with thermal monopoles as well as with lattice sites, where the
local holonomy has close-to-degenerate eigenvalues.
All this points to the correctness of an interpretation in terms of 
(anti)dyon excitations of the KvBLL type and has enabled us to estimate 
corresponding densities and cluster properties.

\noi
{\bf Acknowledgments} \\
%-----------------------
B.V.M. appreciates the support of Humboldt-University
of Berlin where the main part of the work was done.
He also  has been supported by the grants RFBR 13-02-01387a, 15-02-07596a.
E.-M.I. and M.M.-P. acknowledge financial support by the
Heisenberg-Landau Program between the German BMBF and BLTP of JINR Dubna.

%-----------------------------------------------------------------------------
%%\bibliographystyle{apsrev}
%%\bibliography{citations_topology}
%%\end{document}

\end{document}